\documentclass[prd,notitlepage,nofootinbib,preprintnumbers,amssymb,superscriptaddress,twocolumn,eqsecnum]{revtex4-1}
\usepackage{amsfonts,amssymb,amsmath,graphicx,color,bm,ulem}
\usepackage{latexsym,subfigure,ascmac}
\definecolor{ultramarine}{rgb}{0.07, 0.04, 0.56}
\definecolor{cadmiumgreen}{rgb}{0.0, 0.42, 0.24}
\definecolor{indigo(dye)}{rgb}{0.0, 0.25, 0.42}
\usepackage[linktocpage=true]{hyperref}
\hypersetup{
colorlinks=true,
citecolor=ultramarine,
linkcolor=cadmiumgreen,
urlcolor=indigo(dye),
}

\newcommand{\f}[2]{\frac{#1}{#2}}  
\newcommand{\mk}[1]{\left( #1 \right)}

\newcommand{\be}{\begin{equation}}  
\newcommand{\ee}{\end{equation}}

\begin{document}

\title{
Quantum Ostrogradsky theorem
}

\author{Hayato Motohashi}
\thanks{Present address: Division of Liberal Arts, Kogakuin University, 2665-1 Nakano-machi, Hachioji, Tokyo, 192-0015, Japan}
\affiliation{Center for Gravitational Physics, Yukawa Institute for Theoretical Physics, Kyoto University, Kyoto 606-8502, Japan}

\author{Teruaki Suyama}
\affiliation{Department of Physics, Tokyo Institute of Technology, 2-12-1 Ookayama, Meguro-ku, Tokyo 152-8551, Japan}

\begin{abstract}
The Ostrogradsky theorem states that any classical Lagrangian that contains time derivatives
higher than the first order and is nondegenerate with respect to the highest-order derivatives 
leads to an unbounded Hamiltonian which linearly depends on the canonical momenta.
Recently, the original theorem has been generalized to nondegeneracy with respect to non-highest-order derivatives.
These theorems have been playing a central role in construction of sensible higher-derivative theories.
We explore quantization of such nondegenerate theories, and prove that Hamiltonian is still unbounded 
at the level of quantum field theory.
\end{abstract}

\preprint{YITP-20-04}

\maketitle  


\section{Introduction}

The Ostrogradsky theorem states that any Lagrangian which contains more than first-order time derivatives and is non-degenerate with respect to the highest-order derivatives leads to a classical Hamiltonian which is not bounded due to its linear dependence on the canonical momenta~\cite{Ostrogradsky:1850fid, Woodard:2015zca}. It implies that there is in general no stable configuration, which is known as the Ostrogradsky instability. 
Recently, a generalization of the Ostrogradsky theorem has been investigated extensively.  It was proved in~\cite{Motohashi:2014opa} that even though a higher-derivative Lagrangian is degenerate with respect to the highest-order derivatives and hence avoids the original Ostrogradsky theorem, a nondegeneracy with respect to the next-highest order derivatives still makes the Hamiltonian unbounded.  Furthermore, analysis on more general Lagrangian involving multiple kinds of variables with arbitrary higher derivatives was established recently~\cite{Langlois:2015cwa,Motohashi:2016ftl,Motohashi:2017eya,Motohashi:2018pxg}. 
These generalizations of the original Ostrogradsky theorem are powerful and provide severe restriction for constructing consistent higher-derivative theories.

However, the unboundedness of the classical Hamiltonian does not always mean that the corresponding quantum Hamiltonian is also unbounded since the canonical momentum cannot be varied independently of its conjugate variable due to the uncertainty principle. For instance, the Hamiltonian of the hydrogen atom is classically unbounded but bounded from below quantum mechanically. It is then an interesting question to ask if the Ostrogradsky theorem remains true even at the quantum level. 
While the original Ostrogradsky theorem on the highest derivatives was considered at the quantum level in \cite{Raidal:2016wop}, 
there are no explicit and quantitative studies that address this issue for the generalizations of the Ostrogradsky theorem.

In this paper, after reviewing the Ostrogradsky theorem in \S\ref{sec:ost}, we consider more general nondegeneracy structure in \S\ref{sec:3rd}, and prove that the Hamiltonian is unbounded quantum mechanically and has no ground state.  We generalize our argument to quantum field theory in \S\ref{sec:qft}, and further general system with arbitrary higher-order derivatives in \S\ref{sec:gen}.  In \S\ref{sec:con}, we discuss conclusions.

\section{Ostrogradsky theorem}
\label{sec:ost}

In this section, we briefly review the Ostrogradsky theorem for nondegenerate Lagrangians
both at the classical~\cite{Woodard:2015zca} and the quantum~\cite{Raidal:2016wop} level to make our argument self-contained.

\subsection{Classical analysis}

Let us consider a general second-order Lagrangian $L=L(q,{\dot q},{\ddot q})$ of a point particle $q=q(t)$.
Following the Ostrogradsky's method~\cite{Woodard:2015zca},
we define canonical coordinates $q_1=q$, $q_2=\dot q$ and their conjugate momenta as
\be \label{cano}
p_1=\frac{\partial L}{\partial {\dot q}}-\frac{d}{dt} \frac{\partial L}{\partial {\ddot q}}, \quad
p_2=\frac{\partial L}{\partial {\ddot q}}.
\ee
In general, $p_1$ and $p_2$ are functions of $\{ q,{\dot q}, {\ddot q}, {\dddot q} \}$.
The assumption of the Ostrogradsky theorem is that the Lagrangian is nondegenerate with respect to the highest-order derivative, namely, $\partial^2 L / \partial \ddot q^2\ne 0$,
under which the Euler-Lagrange equation is fourth-order differential equation for $q$. 
From the nondegeneracy and the implicit function theorem, we can solve \eqref{cano} for  ${\ddot q}=f(q_1,q_2,p_2), {\dddot q}=g(q_1,q_2,p_1,p_2)$.
Notice that ${\ddot q}$ does not depend on $p_1$.
The Hamiltonian is then given by
\be
H=p_1 {\dot q_1}+p_2 {\dot q_2}-L 
=p_1 q_2 + F(q_1,q_2,p_2),  \label{c-hamiltonian}
\ee
where
$F(q_1,q_2,p_2) \equiv p_2 f(q_1,q_2,p_2) -L(q_1,q_2,f(q_1,q_2,p_2))$.
Hence, the Hamiltonian is linear in $p_1$ and unbounded.
This is the Ostrogradsky theorem at the classical level.

\subsection{Quantum analysis}

In parallel to the quantization of hydrogen atom, one may think a possibility to avoid the Ostrogradsky instability by quantization.
Let us quantize the Hamiltonian (\ref{c-hamiltonian}) by imposing the canonical commutation relations as
\be
[ {\hat q_a}, {\hat p_b} ]=i \delta_{ab},\quad [ {\hat q_a}, {\hat q_b} ]=[ {\hat p_a}, {\hat p_b} ]=0,
\ee
where $a,b$ run $1,2$ and we shall work in the natural unit where $\hbar=1$.
Then the quantum Hamiltonian becomes
\be
{\hat H}={\hat p}_1 {\hat q}_2
+F(\hat q_1,\hat q_2,\hat p_2).
\label{q-hamiltonian}
\ee
This Hamiltonian has ambiguities due to the ordering of the operators.
As it will be clear from the following calculations, our result is not affected by such ambiguities.

If there exists a ground state $|0\rangle$ such that ${\hat H} |0 \rangle =E_0 |0 \rangle$,
we have
$\langle \psi | {\hat H} |\psi \rangle \ge E_0$
for any state $|\psi \rangle$ with the normalization $\langle \psi | \psi \rangle=1$.
Thus, if $\langle \psi | {\hat H} |\psi \rangle$ is not bounded from below, there is no ground state.

Let us consider a state $| \psi \rangle$ defined by
\be
| \psi \rangle =|\phi_1 \rangle |\phi_2\rangle,
\ee
where $|\phi_a \rangle$ is an arbitrary state in the Hilbert space for ${\hat q_a}, {\hat p_a}$
and satisfies the normalization condition $\langle \phi_a | \phi_a \rangle=1$.
The norm of this state is hence unity, $\langle \psi | \psi \rangle=1$.
For this state, the expectation value of ${\hat H}$ is written as
\begin{align}
\langle \psi | {\hat H} |\psi \rangle 
&=\langle \phi_1 | {\hat p_1} | \phi_1 \rangle \langle \phi_2 | {\hat q_2} | \phi_2 \rangle \notag \\
&~~~+\langle \phi_2 | \langle \phi_1 | F({\hat q}_1,{\hat q}_2,{\hat p}_2) |\phi_1 \rangle |\phi_2 \rangle . \label{H-expectation}
\end{align}

Now let us focus on a particular one-parameter family of states $| \phi_1 \rangle$ composed of a superposition of $| x \rangle$, which is the eigenstate of $| \hat q_1 \rangle$, i.e.\ $\hat q_1 | x \rangle = x| x \rangle$,
\be \label{phi1}
| \phi_1 \rangle
= \int dx ~ e^{ikx} g (x) | x \rangle,
\ee
where $k \in \mathbb{R}$ is a parameter, and $g(x)$ is an arbitrary function independent of $k$ and quickly decaying for $|x| \to \infty$ to satisfy the normalization condition $\int dx |g(x)|^2=1$. 
We then obtain
\begin{align}
\langle \phi_1 | {\hat p_1} | \phi_1 \rangle 
&= k + \frac{1}{i} \int dx ~g^* \frac{dg}{dx}, \\
\langle \phi_2 | \langle \phi_1 | F({\hat q}_1,{\hat q}_2,{\hat p}_2) |\phi_1 \rangle |\phi_2 \rangle 
&= \int dx ~|g(x)|^2 \mathcal{F} (x), \label{term-2}
\end{align}
where $\mathcal{F} (x) \equiv \langle \phi_2 | F(x,{\hat q_2},{\hat p_2}) |\phi_2 \rangle$.
It is obvious that Eq.~(\ref{term-2}) is independent of $k$.
Putting these results together, we conclude that $\langle \psi | {\hat H} |\psi \rangle$ can be written as
\be \label{Hp}
\langle \psi | {\hat H} |\psi \rangle = 
C_1 k+C_2,
\ee
where $C_1=\langle \phi_2 | {\hat q_2} | \phi_2 \rangle$ and
$C_2$ depend on $|\phi_2 \rangle$ but not on $k$.
Thus, $\langle \psi | {\hat H} |\psi \rangle$ depends linearly on $k$ and the Hamiltonian is unbounded even at the quantum level.
While \eqref{Hp} is a subset of the full spectrum of the eigenvalues of the Hamiltonian for $| \phi_1 \rangle$ given in \eqref{phi1}, 
its unboundedness means that the full spectrum is also unbounded.

While we have focused on the single-variable second-order Lagrangian~$L=L(q,{\dot q},{\ddot q})$,
the above analysis can be straightforwardly generalized to any multi-variable systems $L=L(q_i,{\dot q_i}, {\ddot q}_i)$
that is nondegenerate with respect to the highest-order derivatives, i.e.\ 
$\det (\partial^2 L / \partial {\ddot q}_i \partial {\ddot q}_j ) \neq 0$, for which the Euler-Lagrange equations consist fourth-order system.

\section{Theory with third-order EOM}
\label{sec:3rd}

In this section, we consider a system which avoids the Ostrogradsky theorem but still has unbounded Hamiltonian.
The unboundedness of the Hamiltonian was shown in \cite{Motohashi:2014opa} at classical level.
Here we generalize it to quantum level.

\subsection{Classical analysis}

To avoid the Ostrogradsky theorem, one can consider a system with $\partial^2 L / \partial {\ddot q}_i \partial {\ddot q}_j=0$.
However, such a system could still suffer from the same type of ghost instability.
Specific form of such a system clarified in~\cite{Motohashi:2014opa} is
\be \label{L3rd} L= \sum^{2N}_{j=1} \ddot q_j f_j(q,\dot q) + c(q,\dot q), \ee 
where $f,c$ are arbitrary functions satisfying $\det C\ne 0$ with an antisymmetric matrix 
$C_{ij} \equiv \f{\partial f_j}{\partial {\dot q}_i}-\f{\partial f_i}{\partial {\dot q}_j}$, 
for which the Euler-Lagrange equations consist third-order system.
It was shown in \cite{Motohashi:2014opa} that the Lagrangian \eqref{L3rd} is the most general one possessing the third-order Euler-Lagrange equations so long as one considers Lagrangian involving at most second-order derivatives. 
As in the previous section, we define canonical coordinates as $q_{1,i}=q_i,~q_{2,i}={\dot q}_i$
and their conjugate momenta as Eq.~(\ref{cano}), i.e.\
\be
p_{1,i}=C_{ij}{\ddot q_j}+\frac{\partial c}{\partial {\dot q_i}}-\frac{\partial f_i}{\partial q_j}{\dot q_j},~~~~
p_{2,i}=f_i (q_1,q_2),
\ee
where the Einstein's summation notation is used.
From the equations for $p_{2,i}$, we immediately see that they provide $2N$ primary constraints
among the canonical variables $(q_{1,i},q_{2,i},p_{1,i},p_{2,i})$ as
\be \label{pricon} \chi_i\equiv p_{2,i}-f_i(q_1,q_2)\approx 0 . \ee
The Hamiltonian is given by
\be \label{H3rd} H = p_{1,j}q_{2,j} - c(q_1,q_2). \ee
Due to $\det C\ne 0$, no secondary constraints arise and the primary constraints
are second class
as the Poisson bracket yields $\{ \chi_i,\chi_j\}_P=C_{ij}$.
As it was argued in \cite{Motohashi:2014opa}, the constraints are not enough to kill the
linear dependence of the Hamiltonian on $p_{1,j}$
and the Ostrogradsky ghosts are still present at the classical level.

\subsection{Quantum analysis}

Below we shall consider the quantization of the system~\eqref{L3rd} and prove that the quantum Hamiltonian is unbounded.
Since $p_{2,i}$ are fixed by the primary constraints~\eqref{pricon}, remaining canonical variables are $(q_{1,i},p_{1,j},q_{2,i})$.
Taking into account the second class constraints, we consider the Dirac bracket
$\{ A, B \}_D \equiv \{ A, B \}_P-\{ A, \chi_k \}_P(C^{-1})_{k\ell} \{ \chi_\ell,B \}_P$.
It is then straightforward to obtain 
\begin{align} \label{dibr}
\{ q_{1,i},p_{1,j} \}_D&=\delta_{ij}, \quad \{ q_{1,i},q_{1,j} \}_D=\{ q_{1,i},q_{2,j} \}_D=0, \notag\\
\{ p_{1,i},p_{1,j} \}_D&=\f{\partial f_k}{\partial q_{1,i}}(C^{-1})_{k\ell} \f{\partial f_\ell}{\partial q_{1,j}}, \\
\{ p_{1,i},q_{2,j} \}_D&=-\f{\partial f_k}{\partial q_{1,i}}(C^{-1})_{kj}, \quad 
\{ q_{2,i},q_{2,j} \}_D=(C^{-1})_{ij}. \notag
\end{align}

Canonical quantization of a constrained system is imposed by promoting the Dirac bracket 
$i\{A, B\}_D$ to the commutation relation $[\hat A, \hat B]$.
We then see that $\hat q_{1,i}$ commute with other variables except $\hat p_{1,j}$.
Hence, the eigenstates of $\hat q_{1,i}$ constitute the basis of the Hilbert space.

We can see that the rest of the basis are the eigenstates of $N$ linear combinations of $\hat q_{2,i}$ as follows.
By virtue of $\det C\ne 0$, 
there exists a linear combination $\bar q_{2,i}= \sum^{2N}_{j=1} A_{ij} q_{2,j}$ such that
the Dirac brackets for the new variables $\bar q_{2,i}$ take a block diagonalized form as
\be \{ {\bar q}_{2,i}, {\bar q}_{2,j} \}_D = 
\begin{pmatrix}  
\sigma & &  \\
& \sigma & \\
& & \ddots
\end{pmatrix}, \quad
\sigma = 
\begin{pmatrix}  
0 & 1 \\
-1 & 0
\end{pmatrix}.
\ee
By using a canonical transformation to redefine $z_n=\bar q_{2, 2n-1}$ and 
$p_{zn}=\bar q_{2, 2n}$ for $n=1,\cdots N$, we have $\{ z_n,p_{zm} \}_D= \delta_{nm}$ and $\{ z_n,z_m \}_D=\{ p_{zn},p_{zm}\}_D=0$.
Here, note that $A_{ij}$ depends on $q_{1,i}$ and $q_{2,i}$ only,
which implies that $q_{2,i}$ is a function of $q_{1,i}$, $z_n$, $p_{zn}$ only,
and does not depend on $p_{1,i}$.
We shall make use of this fact later.

Thus, we have the $3N$ coordinate operators ${\hat q}_{1,i}, {\hat z}_n$ that commute with each other.
The Hilbert space can be then spanned by the eigenstates of these operators: $|x,z \rangle \equiv |x_1 \rangle\cdots|x_{2N} \rangle|z_1 \rangle\cdots |z_N \rangle$,
where ${\hat q}_{1,i} | x_i\rangle = x_i| x_i\rangle$ and ${\hat z}_n | z_n \rangle =z_n | z_n \rangle$.
As in the previous section, let us consider a state 
\be \label{psi}
| \psi \rangle = | \phi_x \rangle | \phi_z \rangle,
\ee
where
\be
| \phi_x \rangle = \int d^{2N}x ~ e^{ik x_1} g ({\vec x}) | {\vec x} \rangle, \quad
| {\vec x} \rangle=|x_1\rangle \cdots | x_{2N} \rangle
\ee
and $| \phi_z \rangle$ is a state spanned by $| z_1 \rangle \cdots | z_N \rangle$.
Here, $k$ is a real parameter and $g({\vec x})$ is an arbitrary function which quickly 
decays as $|{\vec x}|\to\infty$.
The expectation value of the Hamiltonian~\eqref{H3rd} for this state is given by
\be \label{psiHpsi} \langle \psi | \hat H | \psi \rangle =  
\langle \psi | \hat p_{1,j} {\hat q_{2,j}} | \psi \rangle -  \langle \psi | c({\hat q_1},{\hat q_2})| \psi \rangle. \ee 
Using the fact that ${\hat q_2}$ 
does not depend on ${\hat p_1}$,
it is straightforward to show that the second term on the right-hand side is independent of $k$.
In order to determine the first term, let us focus on
the relation $[{\hat q_{1,j}},{\hat p_{1,j}}{\hat q_{2,j}} ]=i{\hat q_{2,j}}$ (no summation over $j$ on the left-hand side).
By sandwiching it by $\langle \phi_z|\langle {\vec x'} |$ and $| {\vec x} \rangle|\phi_z \rangle$, we obtain 
\be (x'_j-x_j) \langle \phi_z|\langle {\vec x'} | \hat p_{1,j} {\hat q_{2,j}} | {\vec x} \rangle|\phi_z \rangle 
= i \langle \phi_z|\langle {\vec x'} | {\hat q_{2,j}} | {\vec x} \rangle|\phi_z \rangle.
\ee
Using again the fact that ${\hat q_2}$ depends on ${\hat q_1}, {\hat z}, {\hat p_z}$ only, we conclude
that the right-hand side is proportional to $\delta ({\vec x}-{\vec x'})$.
Thus, the above equation can be written as
\be (x'_j-x_j) \langle \phi_z|\langle {\vec x'} | \hat p_{1,j} {\hat q_{2,j}} | {\vec x} \rangle|\phi_z \rangle 
= i \delta( {\vec x}-{\vec x'})F({\vec x},{\vec z}).
\ee
From this relation we conclude 
\begin{align} 
\label{pxj} \langle \phi_z|\langle {\vec x'} | \hat p_{1,j} {\hat q_{2,j}} | {\vec x} \rangle|\phi_z \rangle 
=& -i\f{\partial}{\partial x'_j} \delta( {\vec x}-{\vec x'}) F({\vec x},{\vec z}) \nonumber \\
&+\delta( {\vec x}-{\vec x'}) G({\vec x},{\vec z}),
\end{align}
where $G({\vec x},{\vec z})$ is a function that is not determined from the above relation.

Plugging \eqref{pxj} into \eqref{psiHpsi}, we confirm that the Hamiltonian takes the same form as \eqref{Hp},
and hence is unbounded quantum mechanically.

\section{Quantum field theory}
\label{sec:qft}

We can generalize our arguments in \S\ref{sec:ost} and \S\ref{sec:3rd} to quantum field theory.
Consider a single scalar field $\phi=\phi(t,{\vec x})$ in flat spacetime background, and 
a Lagrangian density ${\cal L}={\cal L}(\phi, \partial_\mu \phi, \partial_{\mu} \partial_{\nu}\phi)$ containing up to second derivatives.
First, let us assume the nondegeneracy $\partial^2{\cal L}/\partial\ddot\phi^2\ne 0$.
The canonical variables are $\phi_1=\phi$, $\phi_2={\dot \phi}$.
Then, from the variation of the action, we find the canonical conjugate momenta as
\be \label{pi12}
\pi_1=\frac{\partial {\cal L}}{\partial {\dot \phi}}-\partial_i \left( 
\frac{\partial {\cal L}}{\partial {\dot \phi}_i} \right)
-\partial_t \left( \frac{\partial {\cal L}}{\partial {\ddot \phi}} \right), \quad
\pi_2=\frac{\partial {\cal L}}{\partial {\ddot \phi}}.
\ee
With the nondegeneracy assumption, the implicit function theorem guarantees that we can solve the second equation for ${\ddot \phi}$ and obtain 
${\ddot \phi}=f(\pi_2,\phi_2,\phi_1,\phi_{1,i})$, where $\phi_{1,i}\equiv \partial_i \phi_1$.
Then, the Hamiltonian density is given by
\begin{align} \label{Hqft1}
{\cal H}&=\pi_1 \phi_2+\pi_2 f(\pi_2,\phi_2,\phi_1,\phi_{1,i})\notag\\
&~~~-{\cal L}(\phi_1,\phi_2,\phi_{1,i},f(\pi_2,\phi_2,\phi_1,\phi_{1,i}), \phi_{2,i}, \phi_{1,ij}).
\end{align}
It is clear that ${\cal H}$ linearly depends on $\pi_1$.
Now, we promote the classical fields to the operators in the Schr\"odinger picture and impose the canonical commutation relations
\begin{align}
[ \hat \phi_a ({\vec x}), \hat \pi_b ({\vec y})]&=i \delta_{ab}\delta ({\vec x}-{\vec y}),\notag\\
[ \hat \phi_a ({\vec x}), \hat \phi_b ({\vec y})]&=[ \hat \pi_a ({\vec x}), \hat \pi_b ({\vec y})]=0.
\end{align}
We consider a state given by
\be
| \Psi \rangle = |\Psi_1 \rangle |\Psi_2 \rangle,
\ee
where $|\Psi_a \rangle$ is a state in the Hilbert space for the operators ${\hat \phi_a}, {\hat \pi_a}$. 
Let us consider a set of states parametrized by a real function $k({\vec x})$ as
\be
| \Psi_1 \rangle =\int {\cal D}\psi_1 e^{i \int d^3x ~k({\vec x}) \psi_1 ({\vec x})} G[\psi_1] |\psi_1 \rangle,
\ee
where $|\psi_1 \rangle$ is eigenstate of ${\hat \phi_1}({\vec x})$, namely,
${\hat \phi_1}({\vec x}) |\psi_1 \rangle =\psi_1 ({\vec x}) |\psi_1 \rangle$.
Then, it is straightforward to show that
\be
\langle \Psi | \hat {\cal H} ({\vec x}) | \Psi \rangle
= C_1({\vec x}) k({\vec x})+C_2({\vec x}),
\ee
where $C_1({\vec x}) = \langle \Psi_2| {\hat \phi}_2 | \Psi_2 \rangle$ and 
$C_2({\vec x})$ are independent of $k({\vec x})$.
Thus, the Hamiltonian depends on $k({\vec x})$ linearly and is not bounded from below nor above.

Next, let us consider a generalization of \eqref{L3rd} to field theory, namely,  
\be {\cal L}=\sum_a F^{\mu\nu}_a (\phi,\partial\phi) \partial_\mu\partial_\nu\phi_a +G(\phi,\partial\phi), \ee
with multiple scalar field $\phi_a=\phi_a(t,\vec x)$ and arbitrary functions $F^{\mu\nu}_a$ and $G$.
Here, we assume $\det C\ne 0$, where $C_{ab} \equiv \f{\partial F^{00}_b}{\partial {\dot \phi}_a}-\f{\partial F^{00}_a}{\partial {\dot \phi}_b}$.
We can then show that the Hamiltonian is unbounded as follows.
For the canonical variables $\phi_{1a}=\phi_a$, $\phi_{2a}=\dot\phi_a$, conjugate momenta~\eqref{pi12} read
\be \pi_{1a}=C_{ab}\ddot\phi_b + \cdots, \quad \pi_{2a}=F^{00}_a, \ee
the latter of which is primary constraints.
With the assumption $\det C\ne 0$, we can solve the first equation for $\ddot\phi_a$ to obtain $\ddot\phi_a=f_a(\pi_{1a},\cdots)$.
Also, under $\det C\ne 0$, there are no secondary constraints and the primary constraints are second class.
The Hamiltonian density on the constraint surface is then given by 
\be \label{Hqft2} {\cal H} = \pi_{1a}\phi_{2a} - \mk{ 2 F^{0i}_a\partial_i\phi_{2a} + \cdots + G },  \ee
where $\pi_{1a}$ appears only in the first term linearly.
Then, we can follow the above arguments
and arrive at the same conclusion that even after quantization the Hamiltonian~\eqref{Hqft2} is unbounded from below nor above.

\section{General theories}
\label{sec:gen}

Finally, we consider generalization of our analysis to more general theory.
Since generalization to field theory is straightforward as we see in \S\ref{sec:qft}, here we again focus on analytical mechanics of point particles.

First, let us consider $L=L(\ddot \phi_a, \dot \phi_a, \phi_a, \dot q_i, q_i )$.
We take the canonical variables as $Q_a\equiv \dot \phi_a$, and $\phi_a,q_i$, and their conjugate momenta $P_a,\pi_a,p_i$.
In this case, the following two criteria determine the existence or absence of the Ostrogradsky ghosts~\cite{Motohashi:2016ftl}.
The first criterion is whether a symmetric matrix $K_{ab}\equiv L_{ab}-L_{ai}(L_{ij})^{-1}L_{jb}$ is non/degenerate, where $L_{ab}\equiv\f{\partial^2 L}{\partial \dot Q_a\partial \dot Q_b}$, $L_{ai}\equiv\f{\partial^2 L}{\partial \dot Q_a\partial \dot q_i}$ and $L_{ij}\equiv\f{\partial^2 L}{\partial \dot q_i\partial \dot q_j}$. 
If $\det K\ne 0$ is satisfied, the situation is same as that in \S\ref{sec:ost}, and the Hamiltonian is unbounded, both classically and quantum mechanically.
On the other hand, if $K_{ab}=0$, known as the first degeneracy condition, there exist primary constraints
\be \Xi_a\equiv P_a - F_a (p,Q,q,\phi) \approx 0. \ee

The second criterion is whether an antisymmetric matrix $M_{ab}\equiv \{ \Xi_a,\Xi_b\}_P$ is non/degenerate. 
If $\det M\ne 0$, no secondary constraints arise and the primary constraints are second class.
The on-shell Hamiltonian is then given by 
\be H= F_a\dot Q_a + \pi_a Q_a +p_i\dot q_i -L, \ee
which is unbounded since $\pi_a$ appears only linearly.
Clearly, $\Xi_a$ and $M_{ab}$ correspond to $\chi_i$ and $C_{ij}$ in \S\ref{sec:3rd}.
Further, one can confirm that $Q_a,\phi_a,\pi_a$ have the same structure of the Dirac brackets for $q_{2,i}, q_{1,i}, p_{1,i}$ as in \eqref{dibr}.
Therefore, we can apply the analyses in \S\ref{sec:3rd} and show that the Hamiltonian is still unbounded after canonical quantization.

Further, we can generalize the above argument to general multi-variable Lagrangian containing derivatives higher than second order. 
For simplicity let us focus on a Lagrangian where all the variables have the same $n$-th order derivatives.
Then, there exist $2(n-1)$ degeneracy conditions~\cite{Motohashi:2017eya,Motohashi:2018pxg}. 
Violation of some of the first $n$ degenearcy conditions cause terms linear in canonical momenta, to which we can apply the above analyses.
After imposing the first $n$ degenearcy conditions, linear momentum terms in the Hamiltonian are removed.
However, one still needs to impose remaining $n-2$ degeneracy conditions to obtain a system with number of degrees of freedom same as the number of the variables.
For the case of the quadratic model in \cite{Motohashi:2017eya}, violation of the remaining $n-2$ degeneracy conditions lead to the existence of ``hidden'' ghosts, which show up  only after canonical transformation as terms linear in canonical coordinates in the Hamiltonian.
For these terms, we can apply the above analyses by changing the role of $q$ and $p$, and prove the Hamiltonian is unbounded quantum mechanically.
For more general case, while explicit canonical transformation has not been clarified, it is plausible that violation of remaining $n-2$ degeneracy conditions still causes linear instability, to which we can apply the above analyses.
The logic remains the same for the most general Lagrangian considered in \cite{Motohashi:2018pxg}.

\section{Conclusion}
\label{sec:con}
The original Ostrogradsky theorem focused only on Lagrangian nondegenerate with respect to the highest-order derivatives, and it was known to hold at the level of quantum mechanics.
Recent developments clarified that generalization of the Ostrogradsky theorem is possible, by considering nondegeneracy with respect to non-highest but higher derivatives.
In light of the situation that such a generalization of the Ostrogradsky theorem has been playing a crucial role for construction of sensible higher-derivative theories, 
in this paper we have proved that the unbounded Hamiltonian due to more general nondegeneracy structure 
is still unbounded after quantization.
Our analysis applies to general quantum field theory with arbitrary higher-order derivatives.
It justifies a commonly-adopted way of model building relying on the absence of the Ostrogradsky ghost as a guiding principle at the level of quantum field theory.

\acknowledgments

HM was supported by Japan Society for the Promotion of Science (JSPS) Grants-in-Aid for Scientific Research (KAKENHI) No.\ JP17H06359 and No.\ JP18K13565.
TS was supported by JSPS Grant-in-Aid for Young Scientists (B) No.\ 15K17632,  by the MEXT Grant-in-Aid for Scientific Research on Innovative Areas No.\ 15H05888, No.\ 17H06359, No.\ 18H04338, and No.\ 19K03864.

\bibliography{ref}

\begin{thebibliography}{8}%
\makeatletter
\providecommand \@ifxundefined [1]{%
 \@ifx{#1\undefined}
}%
\providecommand \@ifnum [1]{%
 \ifnum #1\expandafter \@firstoftwo
 \else \expandafter \@secondoftwo
 \fi
}%
\providecommand \@ifx [1]{%
 \ifx #1\expandafter \@firstoftwo
 \else \expandafter \@secondoftwo
 \fi
}%
\providecommand \natexlab [1]{#1}%
\providecommand \enquote  [1]{``#1''}%
\providecommand \bibnamefont  [1]{#1}%
\providecommand \bibfnamefont [1]{#1}%
\providecommand \citenamefont [1]{#1}%
\providecommand \href@noop [0]{\@secondoftwo}%
\providecommand \href [0]{\begingroup \@sanitize@url \@href}%
\providecommand \@href[1]{\@@startlink{#1}\@@href}%
\providecommand \@@href[1]{\endgroup#1\@@endlink}%
\providecommand \@sanitize@url [0]{\catcode `\\12\catcode `\$12\catcode
  `\&12\catcode `\#12\catcode `\^12\catcode `\_12\catcode `\%12\relax}%
\providecommand \@@startlink[1]{}%
\providecommand \@@endlink[0]{}%
\providecommand \url  [0]{\begingroup\@sanitize@url \@url }%
\providecommand \@url [1]{\endgroup\@href {#1}{\urlprefix }}%
\providecommand \urlprefix  [0]{URL }%
\providecommand \Eprint [0]{\href }%
\providecommand \doibase [0]{http://dx.doi.org/}%
\providecommand \selectlanguage [0]{\@gobble}%
\providecommand \bibinfo  [0]{\@secondoftwo}%
\providecommand \bibfield  [0]{\@secondoftwo}%
\providecommand \translation [1]{[#1]}%
\providecommand \BibitemOpen [0]{}%
\providecommand \bibitemStop [0]{}%
\providecommand \bibitemNoStop [0]{.\EOS\space}%
\providecommand \EOS [0]{\spacefactor3000\relax}%
\providecommand \BibitemShut  [1]{\csname bibitem#1\endcsname}%
\let\auto@bib@innerbib\@empty
\bibitem [{\citenamefont {Ostrogradsky}(1850)}]{Ostrogradsky:1850fid}%
  \BibitemOpen
  \bibfield  {author} {\bibinfo {author} {\bibfnamefont {M.}~\bibnamefont
  {Ostrogradsky}},\ }\href
  {http://hdl.handle.net/2027/mdp.39015038710128?urlappend=%3Bseq=405}
  {\bibfield  {journal} {\bibinfo  {journal} {Mem. Acad. St. Petersbourg}\
  }\textbf {\bibinfo {volume} {6}},\ \bibinfo {pages} {385} (\bibinfo {year}
  {1850})}\BibitemShut {NoStop}%
\bibitem [{\citenamefont {Woodard}(2015)}]{Woodard:2015zca}%
  \BibitemOpen
  \bibfield  {author} {\bibinfo {author} {\bibfnamefont {R.~P.}\ \bibnamefont
  {Woodard}},\ }\href {\doibase 10.4249/scholarpedia.32243} {\bibfield
  {journal} {\bibinfo  {journal} {Scholarpedia}\ }\textbf {\bibinfo {volume}
  {10}},\ \bibinfo {pages} {32243} (\bibinfo {year} {2015})},\ \Eprint
  {http://arxiv.org/abs/1506.02210} {arXiv:1506.02210 [hep-th]} \BibitemShut
  {NoStop}%
\bibitem [{\citenamefont {Motohashi}\ and\ \citenamefont
  {Suyama}(2015)}]{Motohashi:2014opa}%
  \BibitemOpen
  \bibfield  {author} {\bibinfo {author} {\bibfnamefont {H.}~\bibnamefont
  {Motohashi}}\ and\ \bibinfo {author} {\bibfnamefont {T.}~\bibnamefont
  {Suyama}},\ }\href {\doibase 10.1103/PhysRevD.91.085009} {\bibfield
  {journal} {\bibinfo  {journal} {Phys. Rev.}\ }\textbf {\bibinfo {volume}
  {D91}},\ \bibinfo {pages} {085009} (\bibinfo {year} {2015})},\ \Eprint
  {http://arxiv.org/abs/1411.3721} {arXiv:1411.3721 [physics.class-ph]}
  \BibitemShut {NoStop}%
\bibitem [{\citenamefont {Langlois}\ and\ \citenamefont
  {Noui}(2016)}]{Langlois:2015cwa}%
  \BibitemOpen
  \bibfield  {author} {\bibinfo {author} {\bibfnamefont {D.}~\bibnamefont
  {Langlois}}\ and\ \bibinfo {author} {\bibfnamefont {K.}~\bibnamefont
  {Noui}},\ }\href {\doibase 10.1088/1475-7516/2016/02/034} {\bibfield
  {journal} {\bibinfo  {journal} {JCAP}\ }\textbf {\bibinfo {volume} {1602}},\
  \bibinfo {pages} {034} (\bibinfo {year} {2016})},\ \Eprint
  {http://arxiv.org/abs/1510.06930} {arXiv:1510.06930 [gr-qc]} \BibitemShut
  {NoStop}%
\bibitem [{\citenamefont {Motohashi}\ \emph {et~al.}(2016)\citenamefont
  {Motohashi}, \citenamefont {Noui}, \citenamefont {Suyama}, \citenamefont
  {Yamaguchi},\ and\ \citenamefont {Langlois}}]{Motohashi:2016ftl}%
  \BibitemOpen
  \bibfield  {author} {\bibinfo {author} {\bibfnamefont {H.}~\bibnamefont
  {Motohashi}}, \bibinfo {author} {\bibfnamefont {K.}~\bibnamefont {Noui}},
  \bibinfo {author} {\bibfnamefont {T.}~\bibnamefont {Suyama}}, \bibinfo
  {author} {\bibfnamefont {M.}~\bibnamefont {Yamaguchi}}, \ and\ \bibinfo
  {author} {\bibfnamefont {D.}~\bibnamefont {Langlois}},\ }\href {\doibase
  10.1088/1475-7516/2016/07/033} {\bibfield  {journal} {\bibinfo  {journal}
  {JCAP}\ }\textbf {\bibinfo {volume} {1607}},\ \bibinfo {pages} {033}
  (\bibinfo {year} {2016})},\ \Eprint {http://arxiv.org/abs/1603.09355}
  {arXiv:1603.09355 [hep-th]} \BibitemShut {NoStop}%
\bibitem [{\citenamefont {Motohashi}\ \emph
  {et~al.}(2018{\natexlab{a}})\citenamefont {Motohashi}, \citenamefont
  {Suyama},\ and\ \citenamefont {Yamaguchi}}]{Motohashi:2017eya}%
  \BibitemOpen
  \bibfield  {author} {\bibinfo {author} {\bibfnamefont {H.}~\bibnamefont
  {Motohashi}}, \bibinfo {author} {\bibfnamefont {T.}~\bibnamefont {Suyama}}, \
  and\ \bibinfo {author} {\bibfnamefont {M.}~\bibnamefont {Yamaguchi}},\ }\href
  {\doibase 10.7566/JPSJ.87.063401} {\bibfield  {journal} {\bibinfo  {journal}
  {J. Phys. Soc. Jap.}\ }\textbf {\bibinfo {volume} {87}},\ \bibinfo {pages}
  {063401} (\bibinfo {year} {2018}{\natexlab{a}})},\ \Eprint
  {http://arxiv.org/abs/1711.08125} {arXiv:1711.08125 [hep-th]} \BibitemShut
  {NoStop}%
\bibitem [{\citenamefont {Motohashi}\ \emph
  {et~al.}(2018{\natexlab{b}})\citenamefont {Motohashi}, \citenamefont
  {Suyama},\ and\ \citenamefont {Yamaguchi}}]{Motohashi:2018pxg}%
  \BibitemOpen
  \bibfield  {author} {\bibinfo {author} {\bibfnamefont {H.}~\bibnamefont
  {Motohashi}}, \bibinfo {author} {\bibfnamefont {T.}~\bibnamefont {Suyama}}, \
  and\ \bibinfo {author} {\bibfnamefont {M.}~\bibnamefont {Yamaguchi}},\ }\href
  {\doibase 10.1007/JHEP06(2018)133} {\bibfield  {journal} {\bibinfo  {journal}
  {JHEP}\ }\textbf {\bibinfo {volume} {06}},\ \bibinfo {pages} {133} (\bibinfo
  {year} {2018}{\natexlab{b}})},\ \Eprint {http://arxiv.org/abs/1804.07990}
  {arXiv:1804.07990 [hep-th]} \BibitemShut {NoStop}%
\bibitem [{\citenamefont {Raidal}\ and\ \citenamefont
  {Veermäe}(2017)}]{Raidal:2016wop}%
  \BibitemOpen
  \bibfield  {author} {\bibinfo {author} {\bibfnamefont {M.}~\bibnamefont
  {Raidal}}\ and\ \bibinfo {author} {\bibfnamefont {H.}~\bibnamefont
  {Veermäe}},\ }\href {\doibase 10.1016/j.nuclphysb.2017.01.024} {\bibfield
  {journal} {\bibinfo  {journal} {Nucl.\ Phys.\ B}\ }\textbf {\bibinfo {volume}
  {916}},\ \bibinfo {pages} {607} (\bibinfo {year} {2017})},\ \Eprint
  {http://arxiv.org/abs/1611.03498} {arXiv:1611.03498 [hep-th]} \BibitemShut
  {NoStop}%
\end{thebibliography}%

\end{document}